\documentclass{llncs}
\usepackage{amssymb}

\newcommand{\Oh}[1]
    {\ensuremath{\mathcal{O}\!\left( {#1} \right)}}
\newcommand{\oh}[1]
    {\ensuremath{o\!\left( {#1} \right)}}

\begin{document}

\title{Counting Colours in\\Compressed Strings}
\author{Travis Gagie\inst{1} \and Juha K\"arkk\"ainen\inst{2}}
\institute{Aalto University, Finland\\
    \email{travis.gagie@aalto.fi}\\
    \mbox{}\\
    University of Helsinki, Finland\\
    \email{juha.karkkainen@cs.helsinki.fi}}
\maketitle

\begin{abstract}
Suppose we are asked to preprocess a string \(s [1..n]\) such that later, given a substring's endpoints, we can quickly count how many distinct characters it contains.  In this paper we give a data structure for this problem that takes \(n H_0 (s) + \Oh{n} + \oh{n H_0 (s)}\) bits, where \(H_0 (s)\) is the 0th-order empirical entropy of $s$, and answers queries in $\Oh{\log^{1 + \epsilon} n}$ time for any constant \(\epsilon > 0\).  We also show how our data structure can be made partially dynamic.
\end{abstract}

\section{Introduction} \label{sec:intro}

Coloured range counting is a well-studied problem with applications in, e.g., computational geometry, database research and bioinformatics.  For this general problem, we are asked to store a set of $n$ coloured points in $\mathbb{R}^d$ such that later, given an axis-aligned box, we can quickly count the number of distinct colours it contains.  Most papers on this problem have focused on \(d \geq 2\) dimensions (see, e.g.,~\cite{KRSV08}); the upper bound for general static one-dimensional coloured range counting has not changed since 1995, when Bozanis, Kitsios, Makris and Tsakalidis~\cite{BKMT95} gave an $\Oh{n}$-word data structure that answers queries in $\Oh{\log n}$ time.  Recently, however, Gagie, Navarro and Puglisi~\cite{GNP10} considered the special case in which the coloured points are the integers \(1, \ldots, n\).  Storing these points is equivalent to storing a string \(s [1..n]\) over an alphabet whose size $\sigma$ is the number of distinct colours, such that later, given a substring's endpoints, we can quickly count how many distinct characters it contains.

Gagie et al. gave a data structure that takes \(n \log \sigma + \Oh{n \log \log n}\) bits and answers queries in $\Oh{\log n}$ time.  (In this paper $\log$ means $\log_2$.)
Their solution is built on work by Muthukrishnan~\cite{Mut02} about coloured range queries in strings.  Muthukrishnan defined \(C [1..n]\) to be the array in which each cell \(C [q]\) stores the largest value \(p < q\) such that \(s [p] = s [q]\) (or 0 if no such $p$ exists).  He observed that \(s [q]\) is the first occurrence of that distinct character in \(s [i..j]\) if and only if \(i \leq q \leq j\) and \(C [q] < i\). Therefore, the number of distinct characters in \(s [i..j]\) is the number of values in \(C [i..j]\) strictly less than $i$.  Gagie et al. noted that, if we store $C$ in a wavelet tree~\cite{GGV03}, which takes \(n \log n + \oh{n \log n}\) bits, then we can count all such values in $\Oh{\log n}$ time; for details, see~\cite{MN07}.  This is already a slight improvement over the bounds we achieve with Bozanis et al.'s data structure~\cite{BKMT95}, but Gagie et al. showed it can be reduced to \(n \log \sigma + \Oh{n \log \log n}\) by modifying the wavelet tree.

In Section~\ref{sec:simple} we describe a simple data structure that achieves essentially the same bound as Gagie et al.'s.  In Section~\ref{sec:multi-size} we extend the ideas from Section~\ref{sec:simple} to build a data structure that takes \(n H_0 (s) + \Oh{n} + \oh{n H_0 (s)}\) bits, where \(H_0 (s) \leq \log \sigma\) is the 0th-order empirical entropy of $s$, and answers queries in $\Oh{\log^{1 + \epsilon} n}$ time for any constant \(\epsilon > 0\).  This may be useful for applications such as tracking the unique visitors to a website, allowing us to count the unique visitors in any given interval.  In Section~\ref{sec:dynamic} we show how our data structure can be made partially dynamic.

\section{Simple Blocking} \label{sec:simple}

In this section we give a simple proof that, using two normal wavelet trees and a straightforward encoding of $C$, we need store only \((1 + \oh{1}) (n \log \sigma + n \log \log n)\) bits to answer queries in $\Oh{\log n}$ time.  Without loss of generality, assume \(\sigma = \oh{n / \log n}\); otherwise, we achieve our desired bound by simply storing $C$ in a single, normal wavelet tree.  Our idea is to break $s$ into blocks of length \(\sigma \log n\) and encode the entry \(C [q]\) differently depending on whether the previous occurrence \(s [p]\) of the character \(s [q]\) is contained in the same block.  If \(s [p]\) is contained in the same block as \(s [q]\), then we write \(C [q]\) as the \(\lceil \log b \rceil\)-bit offset of $p$ within the block; otherwise, we write it as the \(\lceil \log n \rceil\)-bit binary representation of $p$.  Notice that, for each block, there are at most $\sigma$ entries of $C$ encoded as \(\lceil \log n \rceil\)-bit numbers.

We build a bitvector indicating how each entry of $C$ is encoded, which takes \(n + \oh{n}\) bits.  We build one wavelet tree storing all the \(\lceil \log b \rceil\)-bit encodings, which takes at most \(n \log b + \oh{n \log b} = (1 + \oh{1}) (n \log \sigma + n \log \log n)\) bits, and another storing all the \(\lceil \log n \rceil\)-bit encodings, which takes at most \(\sigma \lceil n / b \rceil \log n + \oh{\sigma \lceil n / b \rceil \log n} = n + \oh{n}\) bits.  Notice that, if \(s [q]\) is the first occurrence of that distinct character in \(s [i..j]\) and \(C [q]\) is encoded in \(\lceil \log b \rceil\) bits, then \(s [q]\) must be between \(s [i]\) and the end of the block containing \(s [i]\).  We can count all such characters in \(\Oh{\log b} = \Oh{\log \sigma + \log \log n}\) time using the bitvector and the first wavelet tree.  We can count all the other first occurrences in $\Oh{\log n}$ time using the bitvector and the second wavelet tree.

\begin{theorem} \label{thm:simple}
Given a string \(s [1..n]\), we can build a data structure that takes \((1 + \oh{1}) (n \log \sigma + n \log \log n)\) bits such that later, given a substring's endpoints, in $\Oh{\log n}$ time we can count how many distinct characters it contains.
\end{theorem}

Notice that, if \(\sigma \geq \log n\), then Gagie et al.'s data structure is within a constant factor of being succinct and the data structure we just presented is within a factor of 2 of being succinct.  If \(\sigma < \log n\), then we can store $s$ in a multiary wavelet tree~\cite{FMMN07}, which takes \(n H_0 (s) + \oh{n}\) bits, and answer any query by enumerating the characters in the alphabet and, for each one, using two $\Oh{1}$-time rank queries to see whether it occurs in the given substring.

\begin{corollary} \label{cor:simple}
Given a string \(s [1..n]\), we can build a data structure that takes \(2 n \log \sigma + \oh{n \log \sigma}\) bits such that later, given a substring's endpoints, in $\Oh{\log n}$ time we can count how many distinct characters it contains.
\end{corollary}

\section{Multi-Size Blocking} \label{sec:multi-size}

In this section we extend our idea from the previous section so that, instead of encoding entries of $C$ differently for only two block sizes --- i.e., $\sigma \log n$ and $n$ --- we use many block sizes.  In particular, we use $\Oh{\log \log n / \log (1 + \delta)}$ different block sizes,
\[1, 2^{1 + \delta}, 2^{\max \left( (1 + \delta)^2, 2 \right)}, 2^{\max \left( (1 + \delta)^3, 3 \right)}, 2^{\max \left( (1 + \delta)^4, 4 \right)}, \ldots, n\,,\]
where \(\delta > 0\) is a value we will specify later.  Also, for each block size $b$, we consider $s$ to consist of about \(2 n / b\) evenly overlapping blocks,
\[s [1..b], s [b / 2..3 b / 2], s [b + 1.. 2 b], s [3 b / 2 + 1..5 b / 2], \ldots, s [n - b + 1, n]\,.\]
If \(C [q] = p\) and the smallest block containing both \(s [p]\) and \(s [q]\) has size $b$, then we write \(C [q]\) as the \(\lceil \log b \rceil\)-bit offset of $p$ within the leftmost of the (at most) two blocks of size $b$ containing \(s [q]\).  Notice \(\log b < (1 + \delta) \log (q - p) + 1\); calculation shows that the total size of all the offsets is at most \((1 + \delta) n H_0 (s) + \Oh{n}\) bits.

Let $t$ be a string indicating whether each entry of \(C [q]\) is 0 and, if not, the block size used for it.  We build a multiary wavelet tree~\cite{FMMN07} storing $t$.  Since we can always encode a block size $b$ using $\Oh{\log \log b}$ bits --- even if $\delta$ is very small, thanks to the $\max$ in the definition of the block sizes --- more calculation shows that \(H_0 (t) = \Oh{\log (H_0 (s) + 1)}\).  It follows that, if \(H_0 (s)\) grows without bound as $n$ goes to infinity, then the size of the tree is $\oh{n H_0 (s)}$ bits; otherwise, it is $\Oh{n}$ bits.  Using the tree, in $\Oh{1}$ time we can count all the characters whose first appearance in $s$ is in \(s [i..j]\).

For each block size $b$, we build a wavelet tree storing all the \(\lceil \log b \rceil\)-bit encodings.  By the same calculation as for the offsets, these wavelet trees take a total of \((1 + \delta) n H_0 (s) + \Oh{n} + \oh{n H_0 (s)}\) bits.  Notice that, for any block size $b$, if \(s [q]\) is the first occurrence of that distinct character in \(s [i..j]\) and \(C [q]\) is encoded in \(\lceil \log b \rceil\) bits, then \(s [q]\) must be between \(s [i]\) and the end of the rightmost of the (at most) two blocks of size $b$ containing \(s [i]\).  Using the multiary wavelet tree and the wavelet tree for block size $b$, in $\Oh{\log b}$ time we can count all such characters in the right halves of both the leftmost and the rightmost blocks of size $b$ containing \(s [i]\).  Since the right half of the leftmost block is the left half of the rightmost block, the sum is the total number of such characters.  It follows that we can count all the distinct characters in \(s [i..j]\) in $\Oh{\log n \log \log n / \log (1 + \delta)}$ time.  Choosing \(\delta = 1 / \log \log n\), for example, yields the following theorem:

\begin{theorem} \label{thm:multi-size}
Given a string \(s [1..n]\), we can build a data structure that takes \(n H_0 (s) + \Oh{n} + \oh{n H_0 (s)}\) bits such that later, given a substring's endpoints, in $\Oh{\log n\,(\log \log n)^2}$ time we can count how many distinct characters it contains.
\end{theorem}

A closer analysis shows that the time to count the distinct characters in \(s [i..j]\) is $\Oh{\log (j - i + 1) \log \log n \log \log (j - i + 2)}$.  In a future version of this paper we will improve this bound to $\Oh{\log (j - i + 1) + \min (\log (j - i + 1), \log \log n)^2}$ without increasing our space bound.  As far as we know, no other data structure for coloured range counting has a non-trivial upper bound depending only on the size of the range.

\section{Partial Dynamism} \label{sec:dynamic}

Suppose \(s [i_x]\) and \(s [i_y]\) are the last occurrences of $x$ and $y$ strictly before \(s [j]\), and \(s [k_x]\) and \(s [k_y]\) are their first occurrences strictly after \(s [j]\).  Then to change \(s [j]\) from an $x$ to a $y$, we need only reset \(C [j] = i_y\), \(C [k_x] = i_x\) and \(C [k_y] = j\).  To delete a character from $s$, we replace it with a special null character not in the alphabet (which we search for and exclude when performing queries).  To append a character to $s$, we need only append an entry to $C$.  Assume we have already found all the necessary positions using, e.g., a rank/select data structure for $s$ (although, given some, we can find the others using our data structure from Section~\ref{sec:multi-size}); in this paper we focus on how to update entries of $C$ in our data structure's representation.

M\"akinen and Navarro~\cite{MN08} gave a dynamic data structure that stores a bitvector $v$ of length $n$ in \(n H_0 (v) + \oh{n}\) bits and supports rank, select, insert and delete in $\Oh{\log n}$ time.  Using this dynamic bitvector data structure, they gave an efficient dynamic wavelet tree data structure.  If we simply replace by standard dynamic wavelet trees the two static wavelet trees in our data structure from Theorem~\ref{thm:simple}, then our space bound does not change and it takes $\Oh{\log^2 n}$ time both to count the number of distinct characters in a given substring and to update an entry of $C$.

If we simply replace with standard dynamic wavelet trees all the static wavelet trees (including the multiary wavelet tree) in our data structure from Theorem~\ref{thm:multi-size}, then calculation shows we use \(n H_0 (s) + \Oh{n} + \oh{n (H_0 (s) + \log \log \log n)}\) bits and $\Oh{(\log n \log \log n)^2}$ time both to count the number of distinct characters in a given substring and to update an entry of $C$.  This space bound is $\oh{n \log \log \log n}$ bits larger than the space bound in Theorem~\ref{thm:multi-size} because $t$ --- the string indicating the block size used for each entry of $C$ in Section~\ref{sec:multi-size} --- is over an alphabet of size \(\Oh{\log \log n / \log (1 + \delta)}\).  Therefore, whereas a multiary wavelet tree for $t$ takes \(n H_0 (t) + \oh{n} = \Oh{n} + \oh{n H_0 (s)}\) bits, a standard wavelet tree for $t$ (static or dynamic) takes \(n H_0 (t) + \oh{n \log \log \log n} = \Oh{n} + \oh{n (H_0 (s) + \log \log \log n)}\) bits.  If we use a Huffman-shaped dynamic wavelet tree to store $t$, however, then it takes only \(n (H_0 (t) + 1) + \oh{n (H_0 (t) + 1)} = \Oh{n} + \oh{n H_0 (s)}\) bits.  We will give details in a future version of this paper.

\begin{lemma} \label{lem:dynamic}
We can make our data structure from Theorem~\ref{thm:multi-size} dynamic, without changing its space bound, such that it takes $\Oh{(\log n \log \log n)^2}$ time both to count the number of distinct characters in a given substring and to update an entry of $C$.
\end{lemma}

\begin{theorem} \label{thm:dynamic}
Suppose we have access to a dynamic rank/select data structure storing $s$ such that queries, insertions and deletions all take $\Oh{(\log n \log \log n)^2}$ time.  Then we can build another data structure that takes \(n H_0 (s) + \Oh{n} + \oh{n H_0 (s)}\) bits such that in $\Oh{(\log n \log \log n)^2}$ time we can replace, delete or append a character or, given a substring's endpoints, count how many distinct characters it contains.
\end{theorem}

\section{Acknowledgments}

Many thanks to Veli M\"akinen, Giovanni Manzini, Gonzalo Navarro, Simon Puglisi and Jorma Tarhio, for helpful discussions.

\bibliographystyle{plain}
\bibliography{counting}

\begin{thebibliography}{1}

\bibitem{BKMT95}
P.~Bozanis, N.~Kitsios, C.~Makris, and A.~K. Tsakalidis.
\newblock New upper bounds for generalized intersection searching problems.
\newblock In {\em Proc. ICALP}, pages 464--474, 1995.

\bibitem{FMMN07}
P.~Ferragina, G.~Manzini, V.~M{\"a}kinen, and G.~Navarro.
\newblock Compressed representations of sequences and full-text indexes.
\newblock {\em ACM Transactions on Algorithms}, 3(2), 2007.

\bibitem{GNP10}
T.~Gagie, G.~Navarro, and S.~J. Puglisi.
\newblock Colored range queries and document retrieval.
\newblock In {\em Proc. SPIRE}, pages 67--81, 2010.

\bibitem{GGV03}
R.~Grossi, A.~Gupta, and J.~S. Vitter.
\newblock High-order entropy-compressed text indexes.
\newblock In {\em Proc. SODA}, pages 636--645, 2003.

\bibitem{KRSV08}
H.~Kaplan, N.~Rubin, M.~Sharir, and E.~Verbin.
\newblock Efficient colored orthogonal range counting.
\newblock {\em SIAM Journal on Computing}, 38(3):982--1011, 2008.

\bibitem{MN07}
V.~M{\"a}kinen and G.~Navarro.
\newblock Rank and select revisited and extended.
\newblock {\em Theoretical Computer Science}, 387(3):332--347, 2007.

\bibitem{MN08}
V.~M{\"a}kinen and G.~Navarro.
\newblock Dynamic entropy-compressed sequences and full-text indexes.
\newblock {\em ACM Transactions on Algorithms}, 4(3), 2008.

\bibitem{Mut02}
S.~Muthukrishnan.
\newblock Efficient algorithms for document retrieval problems.
\newblock In {\em Proc. SODA}, pages 657---666, 2002.

\end{thebibliography}

\end{document}